\documentclass[fleqn,twoside]{article}
\usepackage{espcrc2}
\usepackage{graphicx}
\usepackage{subfigure}
\title{Branes in the 5D Abelian Higgs Model}

\author{P. Dimopoulos\address[NTUA]{Physics Department, 
National Technical University, 15780 Zografou Campus,  Athens, Greece}
        \thanks{E-mail: pdimop@central.ntua.gr},
K. Farakos\addressmark[NTUA]\thanks{E-Mail: kfarakos@central.ntua.gr},
C. P. Korthals-Altes\address[CPT]{CNRS--Centre de Physique Th\'eorique, Luminy, BP 907, 13288 Marseille, France}\thanks{E-Mail: altes@cpt.univ-mrs.fr},
G. Koutsoumbas\addressmark[NTUA]\thanks{E-Mail: kutsubas@central.ntua.gr} 
and S. Nicolis\address[LMPT]{CNRS (UMR 6083)--Laboratoire de Math\'ematiques 
et Physique Th\'eorique, Universit\'e de Tours, Parc Grandmont, 37200 Tours, France}\thanks{Speaker. E-Mail: Stam.Nicolis@phys.univ-tours.fr}
}
       
\begin{document}

\begin{abstract}
 We find 3-brane Higgs and Coulomb phases in the 5D
 Abelian Higgs Model and determine the transition surfaces that separate them
 from the usual bulk phases.   
\vspace{1pc}
\end{abstract}

\maketitle

\section{INTRODUCTION}
Anisotropy has been shown to be  a relevant perturbation for gauge theories\cite{FN}. The coupling to 
fermions allows us to put chiral fermions on the lattice in either the overlap or the domain wall variants\cite{Hernandez}. The continuum limit may be 
taken, since a new phase of ``layers'', or 3-branes, 	appears\cite{ANP} and the transition
from this phase to the 5D confining phase turns out to be second order\cite{HKAN}.

In the past year we have studied what happens, when scalar fields are included
\cite{DFKAKN,DFN}. We have found that an additional ``layered'' phase appears. 
It consists of a stack of 3-branes, one lattice spacing apart, that are in 
the Higgs phase and the fields are confined on each layer. 
The phase diagram comprises thus of five phases, three bulk and two layered:
a confining phase  (${\sf S}$), a bulk Higgs phase (${\sf H}_5$) a bulk 
Coulomb phase (${\sf C}_5$), a layered Coulomb phase (${\sf C}_4$) and 
a layered Higgs phase (${\sf H}_4$). 
We obtained the phase diagram by Monte Carlo simulations and mean field theory
calculations. 
Indeed the phase of ``layers'' is found also 
as a solution of the mean field (i.e. classical) equations of motion.

\section{THE MODEL}
We use the standard, compact, Abelian Higgs action, with provision made for 
different couplings along a single ($\hat 5$) direction from the other four.
\begin{equation}
\begin{array}{l}
S= \beta_{g} \sum_x\sum_{1 \le \mu<\nu \le 4}(1-\cos F_{\mu
\nu}(x))\\
+\beta_g^{\prime}\sum_x\sum_{1 \le \mu \le 4}(1-\cos F_{\mu
5}(x))\\
+\beta_{h}\sum _{x} {\rm Re} [4 \varphi^{*}(x)\varphi (x)
- \sum_{1 \le \mu \le 4} \varphi^{*}(x)U_{\hat \mu}(x) \varphi (x+\hat
\mu)]\\
+\beta_{h}^{\prime} \sum _{x} {\rm Re} [\varphi^{*}(x)\varphi (x)
- \varphi^{*}(x)U_{\hat 5}(x) \varphi (x+\hat 5)]\\
+\sum _{x}[(1-2\beta_{R}-4 \beta_{h}- \beta_{h}^{\prime})\varphi^{*}(x)\varphi
(x)\\
+\beta_{R}(\varphi ^*(x)\varphi (x))^2]\\
\end{array}
\label{compactaction}
\end{equation}
The order parameters we will use are the expectation values of the 
 plaquette in the bulk $P_S$ and the 
plaquette in the transverse direction,  $P_T$ defined by 
\begin{equation}
\begin{array}{l}
P_S=\left\langle\frac{1}{6 N^5} \sum_x
\sum_{1 \le \mu<\nu \le 4} \cos F_{\mu \nu}(x)\right\rangle\\
\\
P_T=\left\langle\frac{1}{4 N^5}
\sum_x \sum_{1 \le \mu \le 4} \cos F_{\mu 5}(x)\right\rangle \\
\end{array}
\end{equation}
and the susceptibility of the link in the bulk 
\begin{equation}
\begin{array}{l}
L_S=\frac{1}{4 N^5} \sum_x
\sum_{1 \le \mu \le 4} \cos(\chi(x+\hat \mu) +A_{\hat \mu}(x)-\chi(x))\\
\\
{\cal S}(L_S)=N^5\left(\left\langle L_S^2\right\rangle-\left\langle L_S\right\rangle^2\right)\\
\end{array}
\end{equation}
where $\chi(x)$ is the phase of the Higgs field, $\varphi(x)=\rho(x)\exp({\mathrm{i}}\chi(x))$ and $U_\mu(x)=\exp({\mathrm{i}}A_\mu (x))$. 

\begin{figure}[!h]
\includegraphics[scale=0.6]{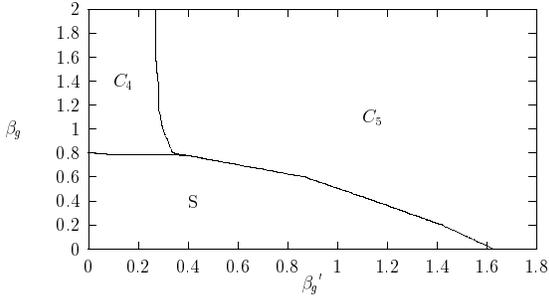}
\caption[]{Phase diagram of the 5D, anisotropic, compact $U(1)$ theory.}
\label{u1phasediag}
\end{figure}

\begin{figure}[!h]
\includegraphics[scale=0.3]{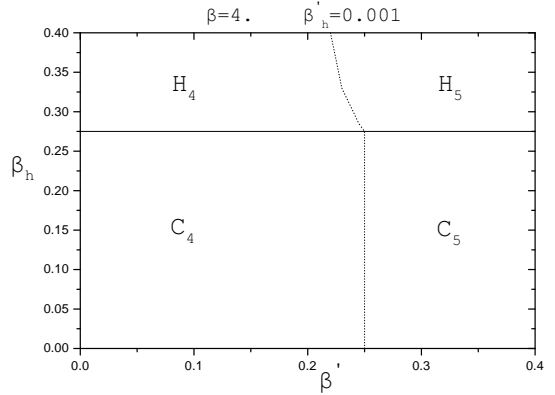}
\caption[]{Phase diagram snapshot for $\beta_g=4$, ${\beta_h}'=0.001$ and $\beta_R=0.01$. Here $\beta'\equiv {\beta_g}'$.}
\label{u1higgsphasediagw}
\end{figure}

\begin{figure}[!h]
\includegraphics[scale=0.3]{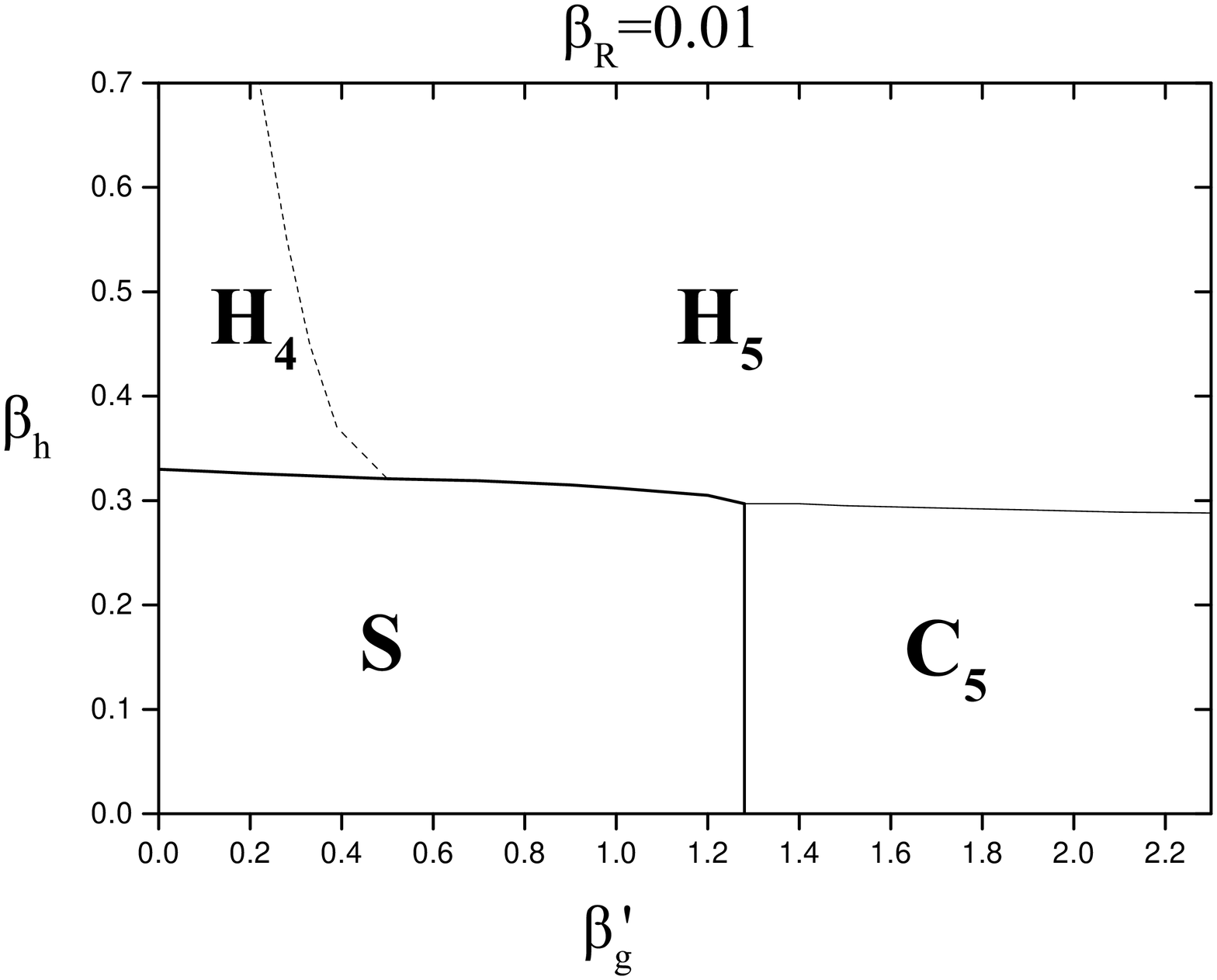}
\caption[]{Phase diagram snapshot for $\beta_g=0.5$, ${\beta_h}'=0.001$.}
\label{u1higgsphasediags}
\end{figure}

\begin{figure}[!ht]
\subfigure[]{\includegraphics[scale=0.2]{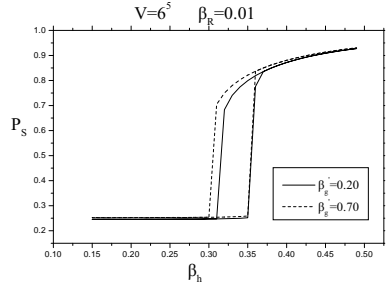}}
\subfigure[]{\includegraphics[scale=0.2]{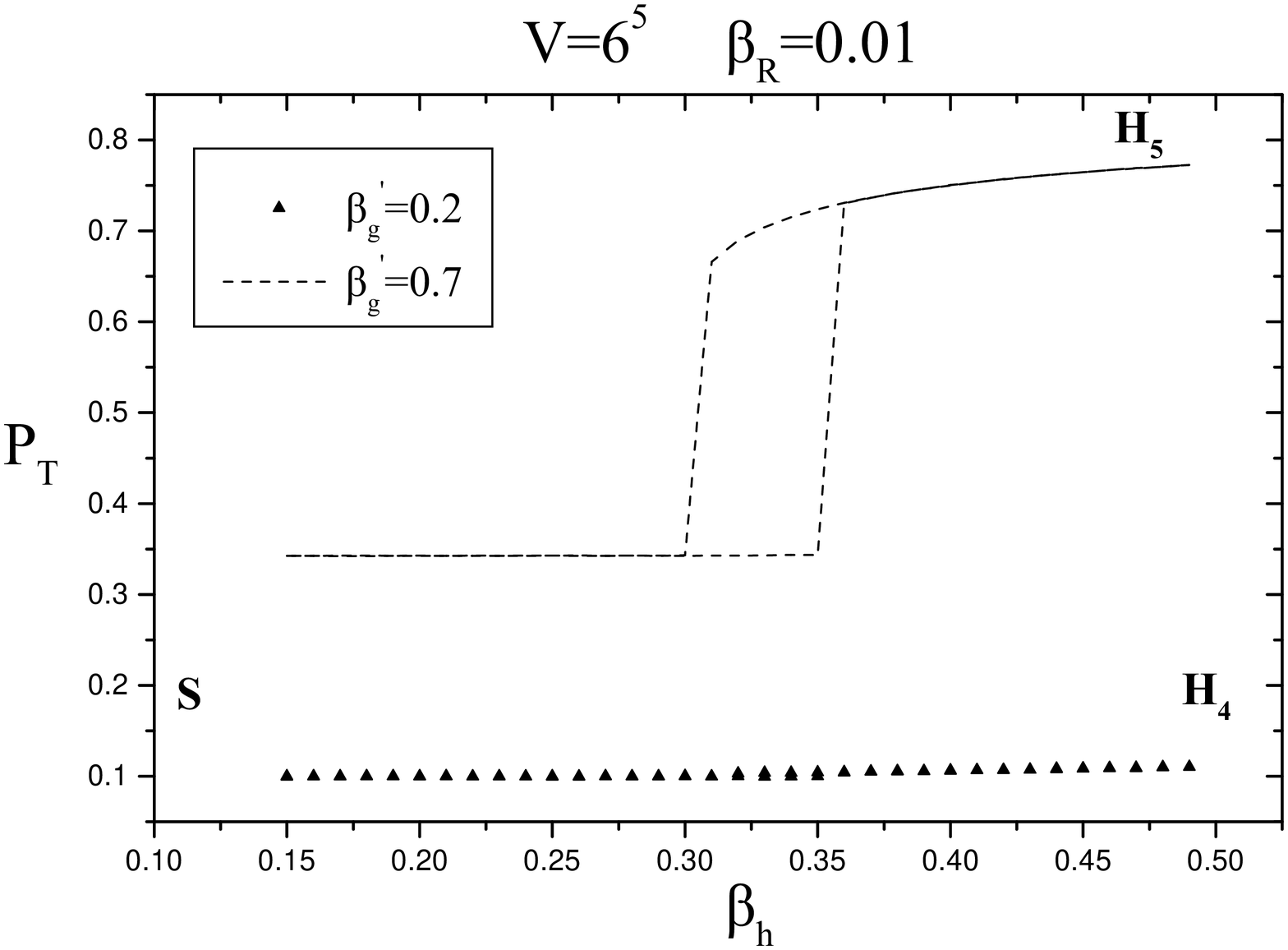}}
\caption[]{Hysteresis loops for the bulk ($P_S$) and transverse ($P_T$)
plaquette expectation values across the transitions from the bulk confining phase 
to the layered Higgs (${\sf H}_4$) and bulk Higgs (${\sf H}_5$) phase.}
\label{Plaquettes}
\end{figure}

\begin{figure}[!h]
\includegraphics[scale=0.3]{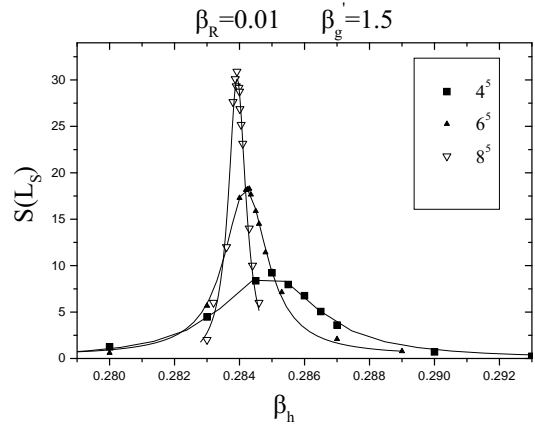}
\caption[]{Susceptibility of the  link in the bulk, ${\cal S}(L_S)$, vs. ${\beta_h}$ for $\beta_g=0.5$, ${\beta_g}'=1.5$, ${\beta_h}'=0.001$ and $\beta_R=0.01$.}
\label{suscept}
\end{figure}

The phase diagram of the 5D, anisotropic, compact $U(1)$ theory is, displayed, for reference purposes,  
in fig.~\ref{u1phasediag}\cite{FN,HKAN}. Including the Higgs 
adds two more dimensions. We fix ${\beta_h}'=0.001$ and $\beta_R,\beta_g$ 
and vary $\beta_h$ and ${\beta_g}'$. For $\beta_g=4$ (weak 4d gauge coupling)\cite{DFKAKN} we find the 
following ``snapshot'' cf. fig.~\ref{u1higgsphasediagw}, while for $\beta_g=0.5$
(strong 4d gauge coupling)\cite{DFN} we find the snapshot in fig.~\ref{u1higgsphasediags}. As expected, at strong coupling the ${\sf C}_4$ phase is no longer there. 
For generic values of the Higgs parameters, $\beta_h$ and ${\beta_h}'$we find 
that the phase transitions are 1st order: we display typical hysteresis loops
for the ``bulk'' and ``transverse'' plaquettes, $P_S$ and $P_T$ in figs.
~\ref{Plaquettes} for the transitions from the 
bulk confining phase ${\sf S}$ to the Higgs phases ${\sf H}_4$ and ${\sf H}_5$.
However, we also find hints of 
continuous transitions, for some subsets of parameter values--cf. the 
susceptibility in fig.~\ref{suscept}. 
\section{CONCLUSIONS-PERSPECTIVES}
We have found 3-brane configurations in the 5D, anisotropic, Abelian 
Higgs model. They may be in either the Coulomb or the Higgs phases. 
The fields in these configurations are {\em confined} on these layers.
This confinement is not put in by hand--it is the defining characteristic
of the layered phase(s).
The transitions between these layered phases and the usual bulk phases are, 
generically, 1st order; however it is also possible to find subsets in
the space of parameter values, that lead to second order transitions. This 
points to the possibility of new, strongly coupled, 
continuum theories\cite{ambjorn}, whose elucidation is of major interest.   
It is to be noted that Yang-Mills theories should exhibit $m$-brane configurations with  $m>3$, since they have a Coulomb phase in more than four dimensions.
Another alternative, relevant for four dimensions,  would be the partial breaking of the Yang-Mills gauge group, that leaves  a $U(1)$ factor in the residual gauge group.  

{\bf Acknowledgements:} P. D., K. F. and G. K. acknowledge support from 
TMR Project ````Finite Temperature
Phase Transitions in Particle Physics", EU contract
FMRX-CT97-0122.


\begin{thebibliography}{9}
\bibitem{FN} Y. K. Fu and H. B. Nielsen, {\sl Nucl. Phys.} {\bf B236} (1984) 167; %%CITATION = NUPHA,B236,167 
{\sl Nucl. Phys. } {\bf B254} (1985) 127. %%CITATION = NUPHA,B254,127
\bibitem{Hernandez} P. Hern\'andez, these Proceedings;
Y. Kikukawa, these Proceedings.
\bibitem{ANP} C. P. Korthals-Altes, S. Nicolis and J. Prades,
{\sl Phys. Lett. } {\bf B316} (1993) 339, [{\tt hep-lat/9306017}]. %%CITATION = HEP-LAT 9306017
\bibitem{HKAN} A. Hulsebos, C. P. Korthals-Altes and S. Nicolis, 
{\sl Nucl. Phys.} {\bf B450} (1995) 437, [{\tt hep-th/9406003}]. %%CITATION = HEP-TH 9406003
A. Hulsebos, {\sl Nucl. Phys. Proc. Suppl.} {\bf 42} (1995) 618, [{\tt hep-lat/9412031}]; %%CITATION = HEP-LAT 9412031
P. Dimopoulos, K. Farakos, A. Kehagias and G. Koutsoumbas, 
[{\tt hep-th/0007079}], {\sl Nucl. Phys. } {\bf B} {\em in press} %%CITATION = HEP-TH 0007079 
\bibitem{DFKAKN} P. Dimopoulos, K. Farakos, C. P. Korthals-Altes, G. Koutsoumbas and S. Nicolis, {\sl J. High Energy Phys.} {\bf 02(2001)005} [{\tt hep-lat/0012028}]. %%CITATION = HEP-LAT 0012028 

\bibitem{DFN} P. Dimopoulos, K. Farakos and S. Nicolis, 
{\em Multi-Layer Structure in the Strongly Coupled 5D Abelian Higgs Model}, 
[{\tt hep-lat/0105014}]; %%CITATION = HEP-LAT 0105014
cf. also P. Dimopoulos, K. Farakos and S. Nicolis, these Proceedings.

\bibitem{ambjorn} J. Ambj\o rn, D. Espriu and N. Sasakura, 
{\sl Mod. Phys. Lett. } {\bf A12} (1997) 2665, [{\tt hep-lat/9707095}]. %%CITATION = HEP-LAT 9707095
\end{thebibliography}
\end{document}